\documentclass{ws-ijmpcs}

\newcommand{\ee}{e^{+}e^{-}}

\newcommand{\pipi}{\pi^{+}\pi^{-}}

\newcommand{\dbartokthreepi}{\bar{D}^{0}\rightarrow K^{+}\pi^{-}\pi^{+}\pi^{-}}

\newcommand{\ra}{\rightarrow}

\newcommand{\Yone}{\Upsilon(1S)}
\newcommand{\Ytwo}{\Upsilon(2S)}

\newcommand{\Yn}{\Upsilon(nS)}

\begin{document}

\markboth{Sookyung Choi} 
{Gyeongsang National University} 

%
\catchline{}{}{}{}{}
%

\title{HADRON AND QUARKMONIUM EXOTICA } 

\author{Sookyung Choi}  

\address{Department of Physics, Gyeongsang National University \\ 
Jinju daero 501, Jinju, 660-773, Republic of KOREA \\
schoi@gsnu.ac.kr\\ } 



\maketitle

\begin{history}
\received{Day November 2013}
\revised{Day Month Year}
\end{history}

\begin{abstract}
A number of charmonium-(bottmonium-)like states have been observed 
in $B$-factory experiments.  Recently the BESIII experiment has joined this search 
with a unique data sample collected at the different center of mass energies
ranging from 3.9 GeV to 4.42 GeV in which they found new charmonium-like states
with non-zero electric charge.
We review the status of experimental searchs for quarkomonium-like states 
and also other types of non-$q\bar{q}$ meson or non-$qqq$ baryons that are
predicted by QCD-motivated models.  We mainly focus on results from the 
$B$-factories and BESIII.

\end{abstract}

\ccode{PACS numbers:}

\section{Introduction}	

Mesons and baryons are formed from color-singlet combinations of quarks. 
All non-$q\bar{q}$ mesons and non-$qqq$ baryons predicted 
by QCD-motivated models are expected to be in color-singlet multi-quark combinations.

Color-singlet multiquark states can be formed from colored diquarks and diantiquarks: 
a pentaquark is predicted to contain two diquarks and one antiquark, the $H$-dibaryon
contains three diquarks, and tetraquark mesons contain a diquark and diantiquark. 
Multiquark states can also be formed by molecules of two (or more) color-singlet
$q\bar{q}$ mesons and/or $qqq$ baryons that are bound together by the exchange of
virtual $\pi$ $\sigma$, $\omega$, etc. mesons.  In this way, Pentaquarks, the $H$-dibaryon,
baryonium and tetraquark mesons can be formed.  If these exist, where are they ?

\section{Pentaquark}

A pentaquark is a predicted baryonic bound state of four quarks and one antiquark.
In the 2003, two experiments, LEPS and CLAS, reported the existence of a positive
strangeness baryon that they called the $\Theta^{+}$. After that, a number of
experiments reported the observation of confirming signals and even other new types
of pentaquarks. But later, high-statistices experiments\cite{pentaq1,pentaq2}
with higher statistics could not reproduce these signals suggesting that 
the LEPS-CLAS results were due to statistical effects rather than a real resonance.
Therefore the existence of the pentaquark is still an open question.
\section{The $H$-dibaryon}

The possibility of a six-quark state was first proposed by R.L.~Jaffe in 1977.\cite{jaffe}
He dubbed it the $H$-dibaryon, and prdicted it to have a strangeness=$-$2 and baryon number=$+$2.
Such a state could
possibly be either a tightly bound diquark triplet or a 
hyperon-hyperon molecular type state.
Jaffe's original prediction was $0^+$ di-hyperon with 
the mass that is about 80 MeV below the 2$m_\Lambda$ threshold.
His original specific prediction was ruled out by observation of  
a double-$\Lambda$ hypernuclei events. Especially the ``NAGARA'' event\cite{nagara}
observed in double-$\Lambda$ hypernuclei, $^{6}_{\Lambda\Lambda}$He, which 
greatly narrowed the mass window of the $H$ to be 
$M_{H}>2m_{\Lambda}$-$B_{\Lambda\Lambda}$~MeV, with binding energy
$B_{\Lambda\Lambda}$= 7.25 $\pm$ 0.19$^{+0.18}_{-0.11}$, corresponding to a
lower limit of the $H$ mass to be 2223.7MeV/c$^2$. 

Although Jaffe's original prediction for $B_H \sim $80 MeV has been ruled out,
the theoretical case for an $H$-dibaryon with a mass near 2$m_\Lambda$ remains concrete.
Recent Lattice QCD results indicate the existence of an $H$-dibaryon  
with the mass close to the 2$m_{\Lambda}$ threshold\cite{latticeQCD} and this
motivated a Belle search for the $H$ near the mass of the 2$m_\Lambda$ threshold.
For an $H$ mass below 2$m_\Lambda$ threshold, the $H$ would decay via a $\Delta S=1$
weak-interaction with a number of possible decay channels: $\Lambda p \pi$,  
$\lambda$ n, $\Sigma$ p, and $\Sigma n$.
For an $H$ mass above the $2\Lambda$ threshold and below the mass of $\Xi p$,  
the $H$ would decay stongly into $\Lambda\Lambda$ almost 100\% of the time.

Belle searched for the $H$-dibaryon production in inclusive $\Upsilon(1,2S) \ra H X$;
$H \ra \Lambda p \pi^-$ and $\Lambda\Lambda$ process using data sample 
containing 102 million events of $\Yone$ and 158 million events of 
$\Ytwo$ collected with Belle detector 
operating at the KEKB $\ee$ collider. 
The dominant decay mechanism for the narrow $\Yn$ resonances (n=1,2 and 3) 
is annihilation into three gluons.  
Each gluon materializes as quark-antiquark pairs and, since the $\Upsilon$ states 
are flavor SU(3) singlets, $u\bar{u}$, $d\bar{d}$ and $s\bar{s}$ pairs 
are produced in nearly equal numbers. 
This strangeness-rich high-density quark environment in a limited volume of phase space
is ideal for producing multi-quark hadron states, especially  $S=-$2, $H$-dibaryon. 

\begin{figure}[pb]
\centerline{\includegraphics[width=10.cm]{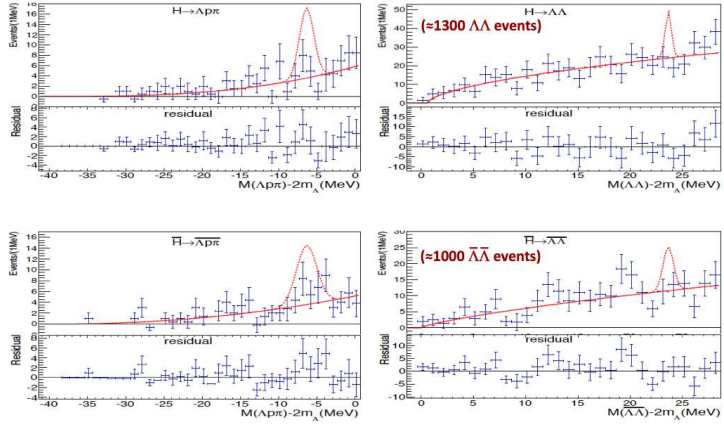}}
\vspace*{8pt}
\caption{ The top and left (right) panel shows the continuum-subtracted 
$M(\Lambda p \pi^{-})$ ($M(\Lambda \Lambda)$) 
distribution (upper) and fit residuals (lower) for the combined $\Yone$ and
$\Ytwo$ data samples.  
The curve shows the results of the background-only fit using an 
ARGUS-like threshold function to model the background. Gaussian fit function
for expected signals is superimposed assuming $\mathcal{B}(\Upsilon \ra HX)=(1/20) \times
\mathcal{B}(\Upsilon \ra \bar{d}X)$.   
The corresponding $M(\bar{\Lambda} \bar{p} \pi^+ )$ ($M(\bar{\Lambda}\bar{\Lambda})$) 
distributions in the bottom left (right) panel.
\label{bongho1}}
\end{figure}

\begin{figure}[pb]
\centerline{\includegraphics[width=6.7cm]{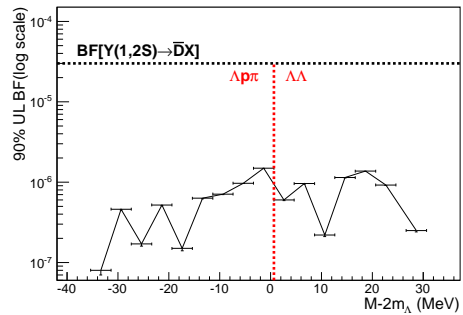}}
\vspace*{8pt}
\caption{ Upper limits at 90\% CL for $\mathcal{B} (\Upsilon(1,2S) \ra HX) 
\mathcal{B} (H\ra f )$ for a narrow ($\Gamma$ = 0) 
$H$-dibaryon vs. $M_{H} - 2m_{\Lambda}$.
The vertical dotted line indicate the $M_{H} $=2$m_{\Lambda}$
threshold. The UL branching fraction below (above) the 2$m_{\Lambda}$ threshold
are for $f = \Lambda p \pi^{-} (f = \Lambda \Lambda)$.
The horizontal dotted line indicates the average PDG 
value for $\mathcal{B} (\Upsilon(1,2S) \ra \bar{D} X)$.
\label{bongho2}}
\end{figure}

These $\Upsilon(1,2S)$ decays are a rich source of 
6-quark antideuteron production, as demonstrated by large branching fraction
of $\sim 3 \times 10^{-5}$ measured by the ARGUS and CLEO experiments.\cite{anti-deutron}   
The physics of near-threshold $S$-wave “Feshbach” resonances ensure
that an $H$-dibaryon in this mass region should be very similar in character 
to a $\Lambda \Lambda$ version of the deuteron. 
Moreover, since the $\Yone$ and $\Ytwo$ are Flavor-SU(3) singlets, 
S$=-$2 counterparts of the antideuteron 
should be produced with sensitivities below or at the rate for antideuteron production. 

The Fig.~\ref{bongho1} shows results from the Belle search
for $H$-dibaryon masses below and above the 2$m_{\Lambda}$ threshold.\cite{bongho-belle}
For masses below threshold (left), the $H$ is searched for as a peak 
in the $\Lambda p \pi^{-}$ invariant mass distribution in inclusive 
$\Upsilon(1,2S) \ra \Lambda p \pi X$ and its charge conjugate decays. 
For masses above threshold (right), the $H$ is searched for in the invariant 
mass distribution of $\Lambda\Lambda$ pairs from $\Upsilon(1,2S) \ra \Lambda \Lambda X$.
The continuum-subtracted $M(\Lambda p \pi^{-}$) (left) and $M(\Lambda\Lambda)$
(right) distributions have no evident $H \ra \Lambda p \pi^{-}$
and $H \ra \Lambda\Lambda$ signals.

Figure~\ref{bongho2} shows the corresponding $M_H -2m_{\Lambda}$ dependent 
upper limits at 90\% CL for
$\mathcal{B}(\Upsilon(1,2S) \ra H X)\mathcal{B}(H \ra \Lambda p \pi^{-}$ and $\Lambda\Lambda)$.
These are all more than an order of magnitude lower than the averaged PDG
value of $\mathcal{B}(\Upsilon(1.2S) \ra \bar{d}X)$, plotted as 
an horizontal dotted line. 
No evidence was found for a signal in any of these modes,
and the most stringent branching-fraction upper limits on $H$-dibaryon
production are determined for masses near the $2m_{\Lambda}$ threshold.
\section{Candidate of baryonium}

Ten years ago, the BESII experiment  reported the observation of a peculiar  
threshold enhancement in the $p \bar{p}$ invariant mass distribution 
in the radiative decay process $J/\psi \ra \gamma p \bar{p}$ using 
58M $J/\psi$ events.\cite{ppbar-steve-bes} 
A fit to the enhancement in the $M(p\bar{p})$ distribution
near the $M=2m_{p}$ mass threshold with an $S$-wave Breit-Wigner function
gave a below threshold peak value of $M=$1859 $^{+3}_{-10}$(stat)~$^{+5}_{-25}$(syst) MeV/$c^2$
and an upper limit on the full width of $\Gamma<$30 MeV at the 90$\%$ CL. 
The enhancement could not be explained by final state interactions 
(FSI) between the $p$ and $\bar{p}$ and there
was no corresponding any known resonance states listed 
in the standard table that could account for this state. 

With (225.2 $\pm$ 2.8) $\times$ 10$^6$ $J/\psi$ events sample collected
with the BESIII detector, a partial wave analysis that included $I=0$ FSI 
between $p$ and $\bar{p}$ gave a mass and upper limit on the width 
of $M=1832^{+19~+18}_{-5~~-17} \pm 19$(model)~MeV/$c^2$
and $\Gamma<$76 MeV at the 90$\%$ CL.\cite{ppbar-bes}
The fitted mass peak is about 40 MeV below the $M=2m_{p}$ mass threshold.   
Also the $\gamma$ polar angle and $p\bar{p}$ decay angle distributions favor 
the  $J^{PC} = 0^{-+}$ quantum number assignment over any other possibility
with statistical significances that are larger than 6.8$\sigma$.

One proposed interpretation for this enhancement is that it is baryonium, 
{\it i.e.} a deuteron-like proton-antiproton bound state produced by standard 
nuclear forces mediated by $\pi$, $\sigma$ and/or $\omega$ exchanges.  
For such a state with mass below the $2m_{p}$ mass threshold, the
$p$ and $\bar{p}$ would annihilate to mesons, while above threshold
it would be expected to ``fall apart'' to $p\bar{p}$ final states almost 100\% 
of the time.  Since $p\bar{p}$ annihilation to $\pipi \eta^{\prime}$ or 3($\pipi$) 
are common and dominant channels for $I=$0, $J^{PC}=0^{-+}$ $p\bar{p}$ annihilations 
searches for the same state in the radiative processes
$J/\psi \ra \gamma \pipi \eta^{\prime}$
and $J/\psi \ra \gamma 3(\pipi)$ were pursued by BESIII with the same  $J/\psi$ data 
sample.

A dramatic peak was seen in the $M(\pipi \eta^{\prime})$ spectrum\cite{threepeaks-bes}
at $M=1837\pm 3 ^{+5}_{-3}$~MeV/$c^2$, very similar to the one measured from $p\bar{p}$ mass, 
also the cosine of the $\gamma$ polar angle distribution is consistent 
with the form of $0^{-+}$.
Although the mass and $J^{pc}$ are consistent with those in the $p\bar{p}$ channel, 
the broad width ($\Gamma = 190 \pm 9 ^{+38}_{-36}$~MeV) is not.
Two more new states at higher masses, dubbed $X(2120)$ and $X(2370)$, are 
also observed in the same $M(\pipi \eta^{\prime})$ mass distribution.   
Recently, BESIII reported the results of a study of the radiative process 
$J/\psi \ra \gamma 3(\pipi)$, where a clear peak near $2m_{p}$ was seen 
with mass $M = 1842.2 \pm 4.2^{+7.1} _{-2.6}$~MeV/$c^2$ and width
$\Gamma = 83 \pm 14 \pm 11$~MeV,\cite{threepipi-bes} consistent with the
the $p\bar{p}$ values. 
Are these peaks observed in the $p\bar{p}$, $\pi\pi \eta ^{\prime}$ and 3$\pipi$ 
mass distribution all from the same state? The 
masses are consistent with each other, but the width measured from $p\bar{p}$ 
is significantly narrower than the width of the $\pipi \eta^{\prime}$ enhancement.
These resonances could come from different sources 
or there may be more than one resonance contributing to the $\pi\pi\eta^{\prime}$ peak.
To clarify all these, further study is needed including the measurement 
of $J^{pc}$ for the $X(1842) \ra 3(\pipi$) signal.

\begin{figure}[pb]
\centerline{\includegraphics[width=7.7cm]{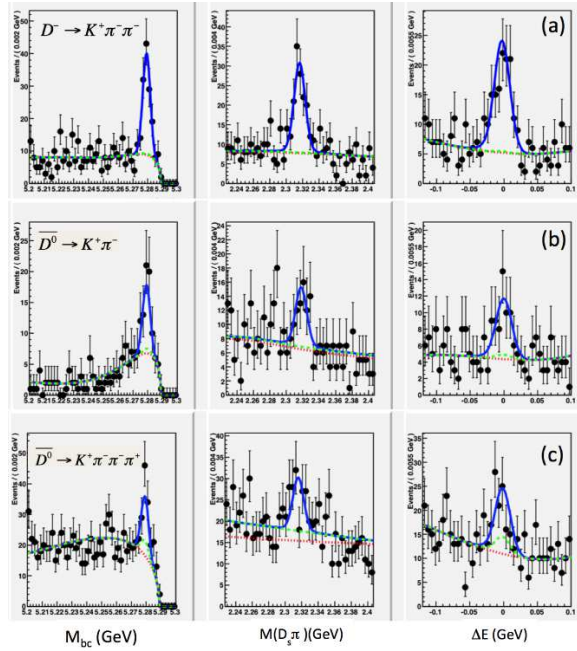}}
\vspace*{8pt}
\caption{The $M_{\rm{bc}}$ (left), $M(D_{s} ^{+} \pi^{0})$ (center) and 
$\Delta E$ (right) distribution for 
a) $B^{0} \ra D^{-} D_{s0} ^{+}$ signal events for the $D^- \ra K^+ \pi^- \pi^-$
sub-decay modes, 
b) $B^{\pm} \ra \bar{D}^0 D_{s0} ^{\pm} ; \dbartokthreepi$ and c) 
$B^{\pm} \ra \bar{D}^{0} D_{s0} ^{\pm} ; \bar{D}^{0} \ra K^+ \pi^- $,
with the results of the fit superimposed.
The events in each distribution are in the signal regions of the two quantities 
not being plotted. 
\label{dc}}
\end{figure}
\section{Doubly charged partner states of $D_{s0} ^+ (2317)$}

The $D_{s0} ^{+} (2317)$, hereafter referred to as the 
$D_{s0} ^{+}$ was first observed by BaBar\cite{ds02317} as a narrow peak 
in the $D_{s0} ^+ \pi^{0}$ mass spectrum produced in $\ee$ annihilation 
to $D_{s0}^+ \pi^{0} + X$ process and subsequently confirmed by CLEO\cite{ds02317-cleo} 
and Belle. 
Its production in $B$ meson decay were also established by both 
Belle\cite{ds02317-belle1} and BaBar\cite{ds02317-2}.
It is generally considered to be the conventional $I(J^{p}) = 0(0^{+})$ p-wave 
$c\bar{s}$ meson, but its mass $M_{D_{s0} ^{+}}= 2317.8 \pm 0.6$~MeV$/c^2$~\cite{pdg2012} 
is very similar as the one $M_{D_0 ^*} = 2318\pm 29$~MeV of 
the non-strange $0^+$ P-wave $D_{0} ^{*}$ state,
in spite of the mass difference between the $s$ and $u$ (or $d$) quarks
which is known as $m_s - m_{u(d)} \sim 100$~MeV.
Furthermore, Potential model 
and lattice QCD 
published prior to its discovery all predicted $0^+$ $P$-wave $c\bar{s}$ 
meson mass to be well above the $m_{D^0} + m_{K^+} = 2358.6$~MeV threshold
with a large partial decay width for $D_{s0} ^+ \ra DK$.
This discrepancy has led to considerable theoretical speculation
that the $D_{s0} ^{+}$ is not a simple $c\bar{s}$ meson,
but instead a four quarks state such as $DK$ molecule or diquark-diantiquark state.

Among these four quarks models, Terasaki's\cite{terasaki-2} assignment 
to $I$=1 iso-triplet four-quark meson
is favored by various existing experimental data. 
If this is the production process of $D_{s0} ^{+}$,
it should have doubly charged $I_{Z}$=1 ($F_{1} ^{++}$)
and neutral $I_{Z}$ = 0 ($F_{1} ^{0}$) partners.\cite{terasaki}
The Ref.~\refcite{terasaki} also predicted that isospin invariance insures 
that the product branching fraction ${\mathcal B}(B^+ \ra D^{-} F_{1} ^{++,0})
{\mathcal B}(F_{1} ^{++,0} \ra D_{s} ^{+} \pi^{+,-})$
will be nearly equal to 
${\mathcal B}(B^{+,0} \ra \bar{D} D_{s0} ^{+})
{\mathcal B}(D_{s0} ^{+} \ra D_{s} ^{+} \pi^{0})$.

Preliminary results from a Belle search uncovered no evidence 
for the predicted state and established a stringent branching fraction limits
to be $\mathcal{B}(B^+ \ra D^- F^{++}(2317))\mathcal{B}(F^{++}(2317) \ra D_{s} ^{+}\pi^{+})$
$< 0.28\times 10^{-4}$ at 90$\%$ CL
that is a factor of about 30 below the predicted level.
This makes possible eliminating the tetra-quark interpretation of 
the $D_{sJ}$ meson family.   

A by-product from this search are measurements of the branching fractions
for the decay processes of $B^+ \ra D_{s0} ^+ \bar{D}^{0}$ and $B^0 \ra D_{s0} ^{+}D^{-}$
with sub-decay $D_{s0} ^+ \ra D_s ^+ \pi^0 ; D_s ^+ \ra K^+K^- \pi^+ $, 
which is significantly improved and used for the above
upper limits calculations for the doubly charged partner state.  
Figure~\ref{dc} shows the $M_{\rm{bc}}$ (left), $M(D_{s} ^{+} \pi^{0})$ (center) and 
$\Delta E$ (right) distribution for $B \ra \bar{D} D_{s0} ^+$ candidate events.
Each distributions are projections of events that are
in the signal regions of other two quantities for modes:
a) $B^{0} \ra D^{-} D_{s0} ^{+} ; D^- \ra K^+ \pi^- \pi^-$, 
b) $B^{\pm} \ra \bar{D}^0 D_{s0} ^{\pm} ; \dbartokthreepi$ and c) 
$B^{\pm} \ra \bar{D}^{0} D_{s0} ^{\pm} ; \bar{D}^{0} \ra K^+ \pi^- $.
The curves in each plot show the results of unbinned three-dimentional
likelihood fit. 
From the fit result, the product branching fraction is determined to be
$\mathcal{B}(B^0 \ra D^{-} D_{s0} ^{+}) 
\mathcal{B}(D_{s0} ^{+} \ra D_{s} ^{+} \pi^{0})$ = $10.0\pm 1.2\pm 1.0 \pm 0.5) 
\times 10^{-4}$ 
and
$\mathcal{B}(B^+ \ra \bar{D}^{0} D_{s0} ^{+}) 
\mathcal{B}(D_{s0} ^{+} \ra D_{s} ^{+} \pi^{0})$ = $(7.8_{-1.2}  ^{+1.3}\pm 1.0 \pm 0.5) 
\times 10^{-4}$,
where the first error is statistical, the second is the 
systematic error, and the third one reflects the errors on the 
PDG world average $D $ and $D_{s} ^{+}$ branching fraction values.
All these measurement used Belle full data containing 
772 million $B\bar{B}$ meson pairs collected at the $\Upsilon(4S)$ resonance.
These results agree well with the average of 
the BaBar and previous Belle measurement with substantial
improvement in precision. 
\section{The $XYZ$ quarkonium-like mesons}

The $XYZ$ states are known as states that decay into final states containing
a $c\bar{c}$ (or $b\bar{b}$) quark pair but do not 
fit into the conventional charmonium (bottomonium) spectrum. 
The $X(3872)$, a key member of this family, was first discovered 
as a peak in the $\pipi J/\psi$ mass by Belle\cite{x3872} 
in the $B \ra \pipi J/\psi K$ decay process and confirmed by 
four different experiments and observed via five different decay channels.
The quantum numbers were unambiguously determined to be $J^{PC}=1^{++}$
by the LHCb experiment\cite{x3872-lhcb} by an analysis of the $B \ra X(3872) K$; 
$X(3872) \ra \pipi J/\psi$ decay chain. This result favors an exotic explanation
of the $X(3872)$ state and disfavors a simple $c\bar{c}$ charmonium assignment.
Recently, one more production channel was observed by BESIII\cite{x3872-bes3},
where a $6.3\sigma$ significance $X(3872)$ signal is reported
in the process $\ee \ra \gamma X(3872)$ using data collected at the four
center of mass energies: $\sqrt{s}= 4.009, 4.229, 4.260\ {\rm and}\ 4.360$~GeV.
Large cross sections of $\gamma X(3872)$ are seen at $\sqrt{s}=$ 4.229 and 4.260,
while cross section at $\sqrt{s}=$ 4.009 and 4.360 GeV were consistent with zero.
This suggests the $X(3872)$ might be produced from the radiative transition 
of the $Y(4260)$ rather than from the $\psi(4040)$ or $Y(4360)$,
but, with the current limited statistics, continuum production cannot be ruled out.

The charged charmonium-like state $Z(4430)^+$, as a controversal member, 
was first observed by Belle\cite{z4430,z4430-1}
as a peak in the $\pi^+ \psi^{\prime}$ mass in $B \ra  K \pi^+ \psi^{\prime}$ decays.
However, the search by BaBar didn't confirm this signal.\cite{z4430-babar}. 
Belle also observed two more charged states, $Z_1 ^+$ and $Z_2 ^+$, that are seen
in the final states $\pi^+\chi_{c1}$ in the exclusive $B\ra K\pi^+ \chi_{c_{1}}$ decays.\cite{z-2}
Recently Belle has reported the measurement of quantum numbers of the $Z(4430)^+$
by performing a full amplitude analysis of $B^0 \ra \psi^{\prime} K^+ \pi^-$
in four dimensions. 
The table~\ref{table_z4430} shows the results from 4-D fit, the preferred
spin-parity hypotheses is $1^+$,
which is favored over $0^-$ by 2.9$\sigma$, but $0^-$
also cannot be ruled out with current statistics.\cite{z4430-2}
\begin{table}[ph]
\tbl{Fit results in the default model. Errors are statistical only.}
{\begin{tabular}{@{}cccccc@{}} \toprule
$J^p$ & $0^-$ & $1^-$  & $1^+$  & $2^-$ &  $2^+$ \\ \colrule
Mass, MeV/$c^2$ & 4479$\pm$16 & 4477$\pm$4    & 4485$\pm$20 & 4478$\pm$22   & 4384$\pm$19 \\
Width, MeV      & 110$\pm$50  & 22$\pm$14 & 200$\pm$40  & 83$\pm$25 & 52$\pm$28 \\
Significance    & 4.5$\sigma$ & 3.6$\sigma$ & 6.4$\sigma$ & 2.2$\sigma$  & 1.8$\sigma$ \\ \botrule
\end{tabular} \label{table_z4430}}
\end{table}
%

The $Y(4260)$ is a $J^{PC}$=$1^{--}$ resonance peak that was 
discovered in the $\ee \ra \pipi J/\psi$ cross section  
by BaBar.\cite{y4260}
It was subsequently confirmed by CLEO\cite{y4260-cleo}, Belle\cite{y4260-belle}
and BESII.\cite{y4260-bes}
A remarkable feature of the $Y(4260)$ is its large partial width of
$\Gamma(Y(4260) \ra \pipi J/\psi)>$1 MeV which is much larger than 
that for typical charmonium.\cite{xhmo} 
This motivated the Belle to investigate whether or not 
there is a corresponding structure in the bottomonium mass region.  
This study found that there is a large anomalous 
cross section for $\ee\ra \pipi \Yn$ (n=1,2,3) for $\ee$ cm 
energies around 10.9 GeV, near the $\Upsilon(5S)$ resonance.\cite{y5s}  
Belle subsequently found that $\Upsilon(5S)$  
decays to $\pipi \Yn$ final states 
are strong sources of charged, bottomonium-like states 
of $Z_{b}(10610)$ and $Z_{b}(10650)$. 
\subsection{The bottomonium-like mesons}

\begin{figure}[pb]
\centerline{\includegraphics[width=8.7cm]{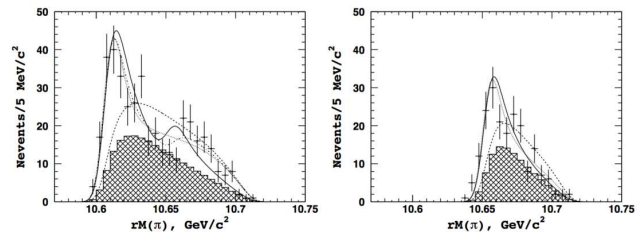}}
\vspace*{8pt}
\caption{$M_{r}(\pi)$ recoil mass distribution for 
$\Upsilon(10860) \rightarrow B\bar{B}^{*} \pi$ (left) 
and $\Upsilon(10860) \rightarrow B^* \bar{B}^* \pi$ (right) candidate events. 
Points with error bars are data, 
the solid line is the results of the fit with the nominal model 
the dashed line - fit to pure non-resonant amplitude, the dotted line - fit 
to a single $Z_b$ state plus a non-resonant amplitude, and the dashed-dotted - 
two $Z_b$ states and a non-resonant 
amplitude. The hatched histogram represents background component normalized to the 
estimated number of background events. 
\label{z1}}
\end{figure}

Two charged bottomonium-like states, the $Z_b(10610)$ and $Z_b(10650)$, 
have been observed in the $\pi^{\pm}\Yn$ and $\pi^{\pm}h_b(mS)$ mass 
at Belle experiment in the decay process $\Upsilon(10860) \ra \pipi \Upsilon(nS)$ (n = 1,2,3) 
and $\pipi h_b(mP)$ (m = 1,2).~\cite{z_b} 
Since they have non-zero electric charge, they must contain 
a minimum of four quarks.

The mass of the	$Z_b (10610)$ is +(2.7 $\pm$ 2.1) MeV above the $M_B + M_B^*$ threshold mass,
while the $Z_b (10650)$ is +(2.0 $\pm$ 1.8) MeV above the 2$M_B^*$.
The proximity to the $B\bar{B}^*$ and $B^*\bar{B}^*$ threshold mass suggest that 
these $Z_b$ states may be a molecular type state formed by two mesons.
This interpretaion can be checked by studying the
$Z_b \ra B\bar{B}^*$ and $Z_b \ra B^*\bar{B}^*$ 
in the $\Upsilon \ra \pi B(B^*) B^*$ three body decay.\cite{z_b2-B*}

Figure~\ref{z1} shows two distinct peaks at the masses of $Z_b(10610)$ 
and $Z_b(10650)$. In this fit the mass was fixed to the earlier measurements. 
This three body decay analysis is extended to 
$\Upsilon(10860) \ra \Upsilon(nS) \pipi$ and $\Upsilon(10860) \ra h_b \pipi$
to measure not only the newly observed $Z_b$ states but also the fractions
of individual contribution to the whole three-body signals.
The relative fractions for $Z_b$ are
$\frac{\mathcal{B}(Z_b(10610) \ra B\bar{B}^*)}
{\mathcal{B}(Z_b(10610) \ra \pi^+ (b\bar{b}))}$ = 6.2 $\pm$ 0.7
and  
$\frac{\mathcal{B}(Z_b(10650) \ra B^*\bar{B}^*)}
{\mathcal{B}(Z_b(10650) \ra \pi^+ (b\bar{b}))}$ = 2.8 $\pm$ 0.4,
here statistical errors added only. 

We started these $\Upsilon(5S)$ mass region search to see $Y(4260)$ analogy in 
$b$ quark sector. Then we found two charged bottomonium-like states which are the smoking guns 
for non $q\bar{q}$ mesons, then the next questions 
are `` are there c-quark versions of $Z_b$’ s ?''. The answer is followed in the next section.
\subsection{The charmonium-like mesons}

\begin{figure}[pb]
\centerline{\includegraphics[width=10.7cm]{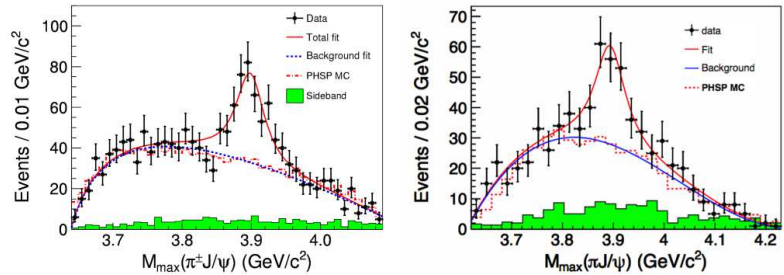}}
\vspace*{8pt}
\caption{
Fit to the $M_{max}(\pi J/\psi)$ distribution from BESIII (left) and Belle (right).
The signal shape is parameterized with same function by using their own 
mass resolution as described in the text,
while background shape is parametrized differently. 
\label{zc3900besbelle}}
\end{figure}

Search for new states were mostly confined to analyses of B-factory data 
including Belle and BaBar, now it has been expanded to include investigations 
of $Y(4260)$ and $Y(4360)$ resonance decays using data from BESIII.
BESIII/BEPCII(Beijing Electron Positron Collider) has been operating 
as a “$Y$ factory” for the past six months, 
collecting large data samples of $\ee$ annihilation events 
at 13 different CM energies from 3.9 GeV to 4.42 GeV,\cite{zcnew-bes1}
especially, at the peaks of the $Y(4230)$ (1090.0 pb$^{-1}$), $Y(4260)$ (826.8 pb$^{-1}$) 
and the $Y$(4360) (~5 $pb^{-1}$) 
resonances, allowing 
in-depth studies for the decay properties of these states. 

Previous studies of these resonances were done at the B-factories 
using the radiative return process, 
which provides rather low luminosities in the charm threshold region.
BESIII studied the process $\ee \ra \pipi J/\psi$ at 
a center of mass energy of 4.26 GeV using 525$pb^{-1}$ data.\cite{zc3900-bes}
The Born cross section is measured to be (62.9 $\pm$ 1.9 $\pm$ 3.7) pb, 
which is consistent with the existing results from the BaBar\cite{y4260-2},
CLEO\cite{y4260-cleo} and Belle\cite{y4260-belle} experiments.
In addition, a surprising structure was observed at the mass of 
(3899 $\pm$ 3.6 $\pm$ 4.9) MeV/$c^2$ and the width of (46 $\pm$ 10 $\pm$
20) MeV in the $\pi^{\pm} J/\psi$ mass spectrum. 
The ratio $R$ of the production rates is obtained to be
$\frac {(\sigma (\ee \ra \pi^{\pm}Z_{c}(3900)^{\mp} \ra \pipi J/\psi ) }
{\sigma (\ee \ra \pipi J/\psi) }$
$= 21.5\pm 3.3(stat) \pm 7.5 (syst)) \% $.

At the same time Belle also measured the cross section for $\ee \ra \pi^+ \pi^- J/\psi$
between 3.8 GeV and 5.5 GeV with 967 fb$^-1$ data sample collected 
by the Belle detector at or near the $\Upsilon(nS)$ (n=1,2,..., 5) 
resonances.\cite{zc3900-belle}
The $Y(4260)$ is observed and its resonant parameters are determined.
In the $Y(4260)$ signal region, the peak state $Z(3900)^{\pm}$ is 
observed with a mass of (3894.5 $\pm$ 6.6 $\pm$ 4.5) MeV/$c^2$ and 
a width of (63 $\pm$ 24 $\pm$ 26) MeV in the $\pi^{\pm} J/\psi$ mass 
spectrum with a statistical significance larger than 5.2$\sigma$. 
This new charged state is refer to as the $Z_{c}(3900)$ or $Z_{c}(3895)$.
Here the ratio of the production rates is obtained from a one-dimentional fit to be 
$\frac {\mathcal{B}(Y(4260)\ra Z(3900)^{\pm} \pi^{\mp}) \mathcal{B}(Z(3900)^{\pm} \ra \pi^{\pm}J/\psi ) } 
 {\mathcal{B}(Y(4260)\ra \pipi J/\psi)  }$ 
= (29.0 $\pm$ 8.9) \%, where the error is statistical only.
For this fit the possible interferences between other amplitides are not included.
This structure couples to
charmonium and has an electric charge, which is suggestive of a state
containing four quarks including $c\bar{c}$ and light quarks $u\bar{d}$.
The Fig.~\ref{zc3900besbelle} shows the fit results from BESIII (left) and 
Belle (right). Unbinned maximum likelihood fit is performed to the
distribution of $M(\pi J/\psi)$, using the larger one of the two combinations.
The same signal shape for both experiments is parameterized as an $S$-wave 
BW function convolved with a Gaussian 
whose mass resolution is fixed at their own MC-estimated values 
depending on detectors.

Reference~\refcite{zcnew-cleo} also reports the observation of this charged state
at a 6$\sigma$ significance level in the analysis of 586 pb$^{-1}$ data taken 
at $\sqrt{s}$=4170 MeV, with the CLEO-c detector at the CESR collider at the 
Cornell University. The search for $Z_c$ was made in the same decay chain 
as the BESIII study, 
$\ee \ra \pi^{\mp} Z_{c} ^{\pm}$, $Z_{c} ^{\pm} \ra \pi^{\pm} J/\psi$, but the data
was taken at $\sqrt{s}$=4170 MeV, on the peak of the $\psi(4160)$
($2^{3}D_{1}$) charmonium state. 
The measured results of the $Z(3900)^{\pm}$ are 
M = (3886 $\pm$ 4(stat) $\pm$ 2(syst) MeV/$c^2$,  
$\Gamma$ = (37 $\pm$ 4(stat) $\pm$ 8(syst)) MeV,
and $R=$ 
$\frac {(\sigma (\ee \ra \pi^{\pm}Z_{c}(3900)^{\mp} \ra \pi^+ \pi^- J/\psi)}
{\sigma (\ee \ra \pi^+ \pi^- J/\psi) } = (32\pm 8 \pm 10) \%$,
These are all in good agreement with the $Z_c(3900)$ signals reported by BESIII and Belle
observed in the decay of $Y(4260)$.
In addtion, the first evidence for the neutral member state, $Z_c ^{0}(3900)$, 
of isospin treplet is also reported which decaying into $\pi^0 J/\psi$ at
a 3.5$\sigma$ significance level.

After this conference, BESIII reported the observation of this 
charged state at one more decay channel: $\ee \ra \pi^{\pm} (D \bar{D}^* )^{\mp}$ 
at the center of mass energy of 4.26 GeV using 525 pb$^{-1}$ data\cite{zcnew-bes3}. 
A distinct charged structure
is observed near the threshold of $m(D)+m(\bar{D}^{ *\pm})$ in the ($D \bar{D}^* )^{\mp}$ 
mass distribution, at $M = (3883.9 \pm 1.5 \pm 4.2) MeV/c^2$, which is 
2$\sigma$ and 1$\sigma$, respectively, below those of the 
$Z_c \ra \pi^{\pm} J/\psi$ peak observed by BESIII and Belle. 
Here the large $Z_c$ signal yield permitted the establishment of the 
$J^P $ quantum number of the $\pi Z_c$ system to
be $1^+$.
%
Assuming the $Z_c(3885) \ra DD^{-̄*}$ signal and 
the $Z_c(3900) \ra \pi J/\psi$ signal are from the
same source, the partial width ratio 
$\frac
{\Gamma(Z_c(3885) \ra D D^{-*})}
{\Gamma(Z_c(3900) \ra \pi^{\pm}J/\psi}$ = 6.2 $\pm$ 1.1 $\pm$ 2.7 is determined.
%

\begin{figure}[pb]
\centerline{\includegraphics[width=4.7cm]{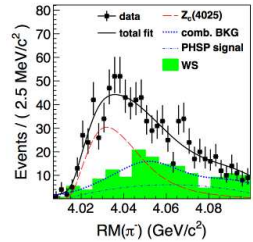}}
\vspace*{8pt}
\caption{Unbinned maximum likelihood fit to the $\pi^-$ recoil mass spectrum in data. 
The $Z_c ^{+}(4025)$ signal shape is taken as the efficiency-weighted BW shape 
convoluted with a dectector resolution function. See the text for a 
detailed description of the other components 
that were used in the fit.
\label{zc4020j}}
\end{figure}

More searches for new charged states at higher mass region were followed 
by exploiting these BESIII scan data sample. 
A charged charmonium-like structure 
is observed near the threshold of $m(D^{*+}) + m(\bar{D}^{*0}) $ in the 
$\pi^{\pm}$ recoil mass spectrum in the process $\ee \ra (D^{*} \bar{D}^{*})^{\pm} \pi^{\mp} $ 
at a center of mass energy of 4.26GeV using a 827 pb$^{-1}$ data.\cite{zcnew-bes2}
Here a partial reconstruction technique is used to identify 
$(D^{*} \bar{D}^{*})^{\pm} \pi^{\mp} $ system, which requires only one 
$\pi^{\mp} $ reconstruction from primary decay, the $D^+ $ from $D^{*+} \ra D^+ \pi^0$,
and at least one soft $\pi^0 $ from either $D^{*+}$ or $\bar{D}^{*0}$.
The Fig.~\ref{zc4020j} shows the $ \pi^{\mp}$ recoil mass spectrum in data
and various components in its fit are described in the Ref.~\refcite{zcnew-bes2}. 
From the fit, the mass and width of the $Z_c ^{+}(4025)$ signal are measured to be 
(4026.3 $\pm$ 2.6(stat) $\pm$ 3.7(syst)) MeV/c$^2$ and 
(24.8 $\pm$ 5.6(stat) $\pm$ 7.7(syst)) MeV, respectively, with the 
statistical significance $> 10 \sigma$.
The ratio $R=$
$\frac
{(\sigma (\ee \ra Z_{c}(4025)^{\pm}\pi^{\mp} \ra (D^* \bar{D}^{*})^{\pm} \pi^{\mp})}
{\sigma (\ee \ra (D^* \bar{D}^{*})^{\pm} \pi^{\mp})}$
is determined to be 0.65 $\pm$ 0.09(stat) $\pm$ 0.06(syst).

\begin{figure}[pb]
\centerline{\includegraphics[width=5.7cm]{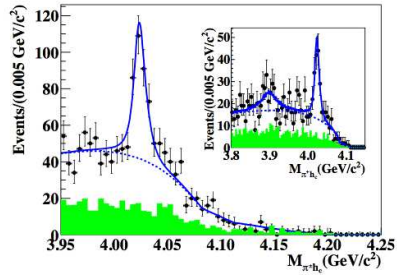}}
\vspace*{8pt}
\caption{ Sum of the simultaneous fits to the $M_{\pi^{\pm}h_{c}}$ distributions
at center-of-mass (CM) energies of 4.23 GeV, 4.26 GeV, and 4.36 GeV.
The inset shows the sum of the simultaneous fit to the $M_{\pi^{\pm}h_{c}}$
distributions at 4.23 GeV and 4.26 GeV with $Z_{c}$(3900) and $Z_{c}$(4020).
\label{zc4020}}
\end{figure}

BESIII has reported searches charged charmonium-like states at the $\pi^{\pm}h_{c}$ 
mass distribution in the process $\ee \ra \pipi h_c$
at 13 different CM energies from 3.9 GeV to 4.42 GeV.\cite{zcnew-bes1}
Here three large statistics CM energies are 4.23 (1090.0), 4.26 (826.8), and 4.36 (544.5 pb$^{-1}$) GeV.  
The $h_c$ is reconstructed via E1 transition $h_c \ra \gamma \eta_c$, and 
the $\eta_c $ is subsequently reconstructed from 16 
different exclusive hadronic final states.
An unbinned maximum likelihood fit is applied to $\pi^{\pm}h_{c}$ mass distribution 
summed over the 16 $\eta_c$ decay modes. Since each data sets at the CM energy 
of 4.23, 4.26 and 4.36 GeV shows similar structures, the same signal function
with common mass and width is used to fit them simultaneously.  
Fig.~\ref{zc4020} shows huge signal named $Z_c(4020)$ with its statistical significance 
greater than 8.9$\sigma$ and fit results give a mass of (4021.8 $\pm$ 1.0 $\pm$ 2.5) MeV/c$^2$ 
and a width of (5.7 $\pm$ 3.4 $\pm$ 1.1) MeV. 
The cross sections are also calculated at each CM energies points.
Adding the $Z_c(3900)$ with mass and width fixed to the measurements of Ref.~\refcite{zc3900-bes}
in the fit results in only a 2.1$\sigma$ significance of it shown in the inset of Fig.~\ref{zc4020}.
The $Z_c(4020)$ agrees within 1.5$\sigma$ of the $Z_c(4025)$ above which observed 
at CM energy 4.26 GeV. 

\section{Summary}

The $J^{PC}$=$1^{--}$ $Y(4260)$ and $\Upsilon(5S)$ have no compelling interpretations.
The $Y(4260)$ ($\Upsilon(5S)$) has 
huge couplings to $\pipi J/\psi$ ($\pipi \Yn$) 
that were not predicted in any model,
and are strong sources of 
charged $Z_c$ ($Z_b$) states observed at $\pi^{\pm} J/\psi$, $\pi^{\pm} h_c$ 
($\pi^{\pm} \Yn$, $\pi^{\pm} h_b$) mass, and 
also with mass near $m(D^{(*)})+m(D^{*})$ ($m(B^{(*)})+m(B^{*})$).

Numerous non‐$q\bar{q}$ mesons not specific to QCD have been found,
such as a baryonium candidate in $J/\psi \ra \gamma p \bar{p}$ at BESII and BESIII
and $XYZ$ mesons that contains $c\bar{c}$ and $b\bar{b}$ pairs together
with light quarks $u\bar{d}$ or $u\bar{u}$.  

QCD motivated spectroscopies predicted by theorists do  not seem to exist. 
No evidence was found for Pentaquarks, and an
$H‐$dibaryon with mass near 2$m_\Lambda$ is excluded at stringent levels.
No hint on $D_{s0} ^{++}$ isospin partner state of $D_{s0} ^+ (2317)$ is observed.
\section*{Acknowledgments}

This work was supported by the Korean Research Foundation via Grant number 2011-0029457.

\end{document}